\documentclass[12pt]{article}
\include{psfig}

\def\be{\begin{equation}}
\def\ee{\end{equation}}
\def\bea{\begin{eqnarray}}
\def\eea{\end{eqnarray}}

\def\NPBPS#1{{\it Nucl. Phys.} {\bf B} ({\it Proc. Suppl.}) {\bf #1}}

\def\PRD#1{{\it Phys. Rev.} {\bf D#1}}
\def\PLB#1{{\it Phys. Lett.} {\bf #1B}}

\def\ms{m_{\rm st}}
\def\str{{\rm str}}

\def\GeV{\mathop{\rm GeV}\nolimits}

\def\tilde{\widetilde}

\def\spose#1{\hbox to 0pt{#1\hss}}
\def\ltapprox{\mathrel{\spose{\lower 3pt\hbox{$\mathchar"218$}}
 \raise 2.0pt\hbox{$\mathchar"13C$}}}
\def\gtapprox{\mathrel{\spose{\lower 3pt\hbox{$\mathchar"218$}}
 \raise 2.0pt\hbox{$\mathchar"13E$}}}
\def\inapprox{\mathrel{\spose{\lower 3pt\hbox{$\mathchar"218$}}
 \raise 2.0pt\hbox{$\mathchar"232$}}}

\begin{document}

\begin{titlepage}
\begin{flushright}
UW/PT 97-18 \\
\today
\end{flushright}
\begin{center}
{\LARGE \bf Enhanced chiral logarithms in \\[.5em]
		partially quenched QCD}
\vspace{1.3cm}

{\large
Stephen R. Sharpe \footnotemark} \\[.5em] 
{\em
Physics Department, Box 351560, University of Washington,
Seattle WA 98195-1560}

\end{center}
\vspace{2cm}

\centerline{\bf ABSTRACT}
\begin{quote}

I discuss the properties of pions in ``partially quenched'' theories,
i.e. those in which the valence and sea quark masses, $m_V$ and $m_S$,
are different.
I point out that for lattice fermions which retain some chiral symmetry
on the lattice, e.g. staggered fermions, the leading order prediction
of the chiral expansion is that the mass of the pion depends only on $m_V$,
and is independent of $m_S$.
This surprising result is shown to receive corrections from
loop effects which are of relative size $m_S \ln m_V$, and which thus
diverge when the valence quark mass vanishes.
Using partially quenched chiral perturbation theory,
I calculate the full one-loop correction to the mass and decay constant
of pions composed of two non-degenerate quarks,
and suggest various combinations for which the prediction is 
independent of the unknown coefficients of the analytic terms in
the chiral Lagrangian.
These results can also be tested with Wilson fermions if one uses
a non-perturbative definition of the quark mass.
\end{quote}

\footnotetext[1]{Email: sharpe@phys.washington.edu}
\vfill
\end{titlepage}

\section{Introduction}\label{sec:intro}

This note is inspired by recent work of the SESAM collaboration,
in which they have studied the light meson spectrum 
with two degenerate flavors of dynamical Wilson quarks \cite{SESAM}.
They have calculated the masses of particles composed not only of sea quarks,
but also of valence quarks with masses 
which can differ from those of the sea quarks.
In the latter case they are studying a ``partially quenched'' theory.
They find that, for their range of quark masses, $M_\pi^2$ can be well 
represented by a linear function of the valence and sea quark masses.
For example, the mass of the non-singlet pion composed of two degenerate
valence quarks of mass $m_V$, calculated with a sea quark mass $m_S$,
takes the form
\be
M_{VV}^2(m_V,m_S)\bigg|_{\rm  Wilson} = c_V m_V + c_S m_S
\,,
\label{eq:SESAM}
\ee
with $c_V \approx c_S$.
The case $m_V=m_S$ corresponds to unquenched pions in two-flavor QCD, 
for which the pion mass vanishes in the chiral limit in the expected way.
What is surprising about Eq.~(\ref{eq:SESAM}) is that
if one works at fixed $m_S$, but extrapolates $m_V\to0$, then
$M_{VV}$ does not vanish. Instead one must go to a
negative valence quark mass, $m_V= - (c_S/c_V) m_S$,
to make the mass of the valence pion vanish.

As explained in Ref.~\cite{SESAM}, this peculiarity is easily understood
in terms of the properties of Wilson fermions.
The point is that chiral symmetry is completely broken by the
lattice regularization, and so the value of hopping parameter $\kappa$ 
at which the valence quark mass vanishes 
depends on the parameters of the theory.
In particular it depends on the sea quark mass, and so one should
define a variable critical hopping parameter, $\kappa_c(m_S)$.
The quark masses in Eq.~(\ref{eq:SESAM}) are, however,
defined as $m=(2\kappa)^{-1}-(2\kappa_c(m_S=0))^{-1}$,
where $\kappa_c(m_S=0)$ is the critical value for which 
the unquenched pion mass vanishes.
The negative quark masses are then an artifact of using 
$\kappa_c(m_S=0)$ instead of $\kappa_c(m_S)$ in the definition of 
$m_V$.\footnote{%
Even though one expects $\kappa_c$ to depend on $m_S$,
the strength of the dependence of found in Ref.~\cite{SESAM}
is surprising, and has interesting implications 
for the extraction of physical quark masses \cite{tanmoy}.}

My main aim in this note is to discuss how the results would change
were one to use fermions for which some remnant of the continuum chiral
symmetry survives discretization. What I have in mind are staggered 
and ``domain-wall'' fermions \cite{domain}.\footnote{%
For the sake of brevity, I will refer only to staggered fermions in the
following, although all such references apply equally to domain-wall fermions.}
In the former case
an axial subgroup of the $SU(4)$ chiral symmetry remains on the lattice,
while in the latter the full chiral symmetry is broken only by exponentially
small corrections. The only property of both types of fermion that I need
is that these symmetries become exact when the lattice 
quark mass vanishes.
If I then assume that the chiral symmetry associated with the valence
quark is broken dynamically, with the formation of a non-zero
condensate $\langle \overline q_V q_V\rangle$, it follows that
there will be a Goldstone pion whose mass vanishes when $m_V=0$.
In other words, the assumption of dynamical breaking of the valence quark
chiral symmetry implies
\begin{equation}
M_{VV}(m_V=0,m_S)=0
\,.
\label{eq:mpivanishes}
\end{equation}
Numerical evidence suggests that
such symmetry breaking does occur for all values of the sea quark mass
(including the quenched case $m_S\to\infty$).

I now add to this the assumption of linearity,
namely that Eq.~(\ref{eq:SESAM}) holds also for staggered fermions.
These two assumptions then imply that $c_S=0$, so that 
\begin{equation}
M_{VV}^2(m_V,m_S)\bigg|_{\rm staggered} = c_V m_V
\,.
\label{eq:stagg}
\end{equation}
In other words, $M_{VV}$ is independent of the sea quark mass,
at least for small $m_S$ where linearity applies.\footnote{%
This result is for fixed bare coupling, $g_0$, because the
``constant'' $c_V$ depends upon $g_0$.}
If correct, this would be a surprising result.
For example,
one could obtain the mass dependence of the physical light pion mass
without the need to work with physical sea quarks.
This seems implausible on physical grounds.
For one thing, the cloud of light mesons surrounding the pion 
depends on $m_S$.
But perhaps such effects are of higher order than
linear in the expansion in the quark masses---indeed,
loop effects in chiral perturbation theory lead to corrections to $M_\pi^2$
proportional to $m_q^2 \ln m_q$.

In fact, I will show that such loop effects are enhanced in the
partially quenched theory. Although the leading term does take the
form of Eq.~(\ref{eq:stagg}), the dominant correction
for small $m_V$ is proportional to $m_V m_S \ln m_V$.
For fixed $m_S$ this correction gets arbitrarily large relative to the
leading order term as $m_V\to 0$.
Thus, as one approaches this limit,
the valence pion mass obtains a significant dependence on $m_S$.
This breakdown of Eq.~(\ref{eq:stagg}) occurs because the assumption
of linearity fails, due to the appearance of non-analytic terms.
The assumption that the valence pion mass vanishes when
$m_V\to0$ remains valid.

My main conclusion is thus that 
one cannot use $M_{VV}$ with unphysical sea quark masses
to give an accurate estimate of the mass of the physical pion.
What one can use $M_{VV}$ for, however, is to provide a sensitive test of 
loop effects predicted for the partially quenched theory.
To this end, I have calculated the one-loop corrections to
both the pion mass and decay constant, as a function of valence and
sea quark masses, using partially quenched chiral perturbation theory
\cite{BGPQ}. In general these predictions depend on unknown constants
multiplying the analytic terms of $O(m_q^2)$, but for
certain combinations the analytic terms cancel.

A similar deviation from
linearity is predicted for $M_{VV}^2$ in fully quenched QCD.
Using quenched chiral perturbation theory \cite{BG,sschlog},
one finds a correction proportional to $m_V m_0^2 \ln m_V$.
Here $m_0$ is a parameter which, in the QCD chiral Lagrangian, gives
the $\eta'$ its mass. There is some numerical evidence supporting this
prediction, but the situation is muddied by the possibility that finite
volume errors mimic the chiral logarithm. For a review, see
Ref.~\cite{sslat96}. The new predictions presented here
provide another way of searching for chiral logarithms,
and thus may help clarify the situation in quenched QCD.
The advantage of partially quenched theories is that,
for reasons explained in the following, the unknown
parameter $m_0$ does not appear in the predictions.

This note is organized as follows. In Sec.~\ref{sec:calc} I
give a brief description of the method of calculation, 
and then present the results for pion
masses and decay constants in Sec.~\ref{sec:res}.
Section~\ref{sec:baryon} contains some general comments on predictions
for partially quenched baryons, and Sec.~\ref{sec:conc} some conclusions.

\section{Calculation}
\label{sec:calc}

Partially quenched chiral perturbation theory has been described in detail
in Ref.~\cite{BGPQ}. I give here a summary of the aspects
relevant to the present calculation.

Consider a theory with two valence quarks, of mass $m_1$ and $m_2$,
and $N\ge1$ unquenched quarks of mass $m_S$.
To cancel internal loops containing the valence quarks one needs
two ghost quarks (commuting quark fields $\tilde q$) 
with masses $m_1$ and $m_2$.
For $N=2$ this is the theory studied in Ref.~\cite{SESAM}.
Collecting all fields into a vector,
\begin{equation}
Q = (q_{V1}, q_{V2}, q_{S1}, q_{S2}, \dots, q_{SN}, 
\tilde q_{V1}, \tilde q_{V2})
\end{equation}
one sees that
the chiral symmetry is the graded group $SU(2+N|2)_L\times SU(2+N|2)_R$.
The chiral Lagrangian consistent with this symmetry is
\bea
{\cal L} &=& {f^2 \over 4}
{\rm str}\left(\partial_\mu\Sigma\partial^\mu\Sigma^\dagger\right)
+ {f^2 \over 4} {\rm str}(\chi \Sigma^\dagger + \Sigma \chi)
	+  \alpha_\Phi \partial_\mu\Phi_0\partial^\mu\Phi_0
		-m_{0}^2\,\Phi_{0}^2, \nonumber
\nonumber \\
	&+& {1 \over 128 \pi^2} \left\{\alpha_4\; 
	\str\left(\partial_\mu\Sigma\partial^\mu\Sigma^\dagger\right)
	\;\str\left(\chi \Sigma^\dagger + \Sigma\chi\right)
	+ 2 \mu \alpha_5 \;
	\str\left(\partial_\mu\Sigma\partial^\mu\Sigma^\dagger
	[\chi\Sigma^\dagger + \Sigma\chi]\right) \right.
\nonumber \\
	&& \mbox{}\qquad\qquad \left.
	+ \alpha_6 \;\str\left(\chi \Sigma^\dagger + \Sigma\chi\right)^2
	+ \alpha_8 \;\str\left(\chi \Sigma^\dagger\chi \Sigma^\dagger
	   + \Sigma\chi\Sigma\chi\right) \right\} + \dots
\,.
\label{eq:chiralL}
\eea
Here $\Sigma=\exp(2i\Phi/f)$ contains all the Goldstone bosons,
including the flavor singlet field\footnote{%
The factor of $\sqrt3$ in $\Phi_0$ 
is chosen so that, if the same normalization
were used in the corresponding QCD chiral Lagrangian,
one would find  $m_{\eta'}^2 = m_0^2/(1+\alpha_\Phi) + O(m_q)$.}
$\Phi_0 = {\rm str}\Phi / \sqrt3$.
The quark masses enter through $\chi= 2\mu M$,
where $M$ is the mass matrix, 
\be
M = {\rm diag}(m_1,m_2,m_S,m_S,\dots,m_S,m_1,m_2)
\,,
\ee
with $N$ entries of $m_S$.
I have only kept those terms in the Lagrangian that will be required
for the following calculations. In particular, I have omitted the
arbitrary function of $\Phi_0$ which can multiply each term.

The terms multiplied by the coefficients $\alpha_i$ are non-leading
in the chiral expansion, since they contain an
additional power of $p^2$ or $m$ compared to the leading order 
terms.\footnote{%
For a full list of these higher order terms see, for example, Ref.~\cite{DGH}.}
They give rise to the corrections to physical quantities which
are analytic in the external momenta and quark masses.
The other source of corrections is loop diagrams involving vertices and
propagators coming from the leading order Lagrangian. These give rise
to the non-analytic ``chiral logarithms'', as well as analytic contributions.
The typical size of both higher order corrections is $p^2/\Lambda_\chi^2$,
where the chiral scale is $\Lambda_\chi=4 \pi f$.

The parameters in the chiral Lagrangian---$f$, $\mu$, $m_0$, $\alpha_\Phi$
and the $\alpha_i$---are not known {\em a priori}.
They are functions of $N$, the number of sea quarks,
and thus are different for QCD ($N=3$) and the $N=2$ case considered in 
Ref.~\cite{SESAM}. 
The most useful predictions of partially quenched
chiral perturbation theory are thus those that depend on as few
of these parameters as possible. 

The calculation of the one-loop corrections to the masses and
decay constants is straightforward. It is very similar
at all stages to the corresponding calculation in quenched QCD,
which has been discussed in Refs.~\cite{BG,sschlog}.
The only significant difference occurs in the propagators of mesons
which have a flavor-singlet component.
Consider, for example, the propagator of the meson with flavor
composition $\bar q_{V1} q_{V1}$.
In quenched QCD this is
\be
G_{11}^{(Q)}(p^2) = {1\over p^2 + M_{11}^2} - 
{(m_0^2 + \alpha_\Phi p^2)/3 \over (p^2 + M_{11}^2)^2}
\,,
\ee
where $M_{11}^2= 2 \mu m_1$ is the leading order mass.
The first term is the usual non-singlet propagator, while the
second arises from ``hairpin'' diagrams in which the quark and antiquark
annihilate. It is the second term which leads to enhanced chiral logarithms.
The corresponding propagator in the partially quenched theory is different
because of the possibility of internal loops of sea quarks.
The result is \cite{BGPQ}
\bea
\lefteqn{G_{11}^{(PQ)}(p^2) - {1\over p^2 + M_{11}^2}}\nonumber\\
&=& -\,
{(m_0^2 + \alpha_\Phi p^2)/3 \over (p^2 + M_{11}^2)^2}
{1\over 1 + (N/3) (m_0^2 + \alpha_\Phi p^2)/(p^2+M_{SS}^2)}
\\
&=& -\, {(m_0^2 + \alpha_\Phi p^2)/3 \over 1 + \alpha_\Phi (N/3)}
\left(
{\tilde M^2 - M_{SS}^2 \over (\tilde M^2 - M_{11}^2)^2} 
\left[{1\over p^2+M_{11}^2} - {1\over p^2+\tilde M^2}\right]
\right.\nonumber\\
&&\mbox{}\left.\qquad\qquad\qquad\qquad +
{M_{SS}^2 - M_{11}^2 \over \tilde M^2 - M_{11}^2} 
{1\over (p^2+M_{11}^2)^2} \right)
\,.
\eea
Here 
\be
\tilde M^2 = {(N/3) m_0^2 + M_{SS}^2  \over 1 + \alpha_\Phi (N/3)}
\ee
is the mass of the singlet ``$\eta'$'' 
meson in the unquenched $SU(N)$ sector of the theory.
Similar results hold for the other flavor-singlet propagators.

The $\eta'$ mass $\tilde M$ does not vanish in the chiral limit.
In the following, I will simplify the calculation by assuming that
the $\eta'$ is a heavy particle, i.e. $\tilde M\approx \Lambda_\chi$,
so that it can be integrated out of the theory.
This is equivalent to assuming that the ratios
$M_{SS}^2/\tilde M^2$, $M_{11}^2/\tilde M^2$, etc. are small.
This is certainly reasonable for $N=3$, for which
we know that $\tilde M\approx M_{\eta',\rm phys}\approx 1\GeV$.
Even for $N=2$, it is a sensible approximation,
since the $\eta'$ will be comparable in mass to the vector mesons,
which we do not include in the chiral Lagrangian.

This assumption leads to two simplifications.
First, integrals involving the $\eta'$ propagator $1/(p^2+\tilde M^2)$
can be dropped. These integrals can be expanded in powers of
$M_{SS}^2/\tilde M^2$, $M_{11}^2/\tilde M^2$, etc.
and their contributions can be absorbed by changing
the coefficients in the chiral Lagrangian.
For the quantities considered here
the effect of $\eta'$ loops is to shift $\mu$, $\alpha_6$
and $\alpha_8$, as I have checked by explicit calculation. 
Since we do not know these parameters {\em a priori},
we lose nothing by dropping the 
contribution from the $\eta'$ propagator.
Indeed, this allows the results for $N=3$ to be matched directly
onto those from the usual QCD chiral Lagrangian, from which the
$\eta'$ has been integrated out.

The second simplification is of the part of
$G_{11}^{(PQ)}$ which remains after the $\eta'$ contribution
has been removed.
In this remainder, we can discard terms suppressed by
powers of $M_{SS}^2/\tilde M^2$, etc.
for these are of the same size as two-loop terms which are not included. 
The propagator then simplifies to
\be
G_{11}^{(PQ)}(p^2) \approx {1\over p^2 + M_{11}^2} - {1\over N}
\left( {1 \over p^2 + M_{11}^2} + 
{M_{SS}^2-M_{11}^2 \over (p^2 + M_{11}^2)^2} \right)
\,.
\ee
Note that the double pole term remains (and is the source of the enhanced
chiral logarithms), but that the unknown parameters
$m_0^2$ and $\alpha_\Phi$ do not appear. 
In the unquenched theory ($M_{SS}=M_{11}$),
the propagator goes over to the usual form, with only a single pole,
and with the $1/N$ term projecting against the $\eta'$.
The corresponding form for the off-diagonal propagator between
a meson of composition $\bar q_{V1} q_{V1}$ and $\bar q_{V2} q_{V2}$ is
\be
G_{12}^{(PQ)}(p^2) \approx  - {1\over N}
\left( 
{M_{SS}^2-M_{11}^2 \over M_{22}^2 - M_{11}^2}
{1 \over p^2 + M_{11}^2}  + [1\leftrightarrow2]\right)
\,.
\ee

\section{Results}
\label{sec:res}

I have calculated the complete one-loop correction to the mass and
decay constant of a pion with composition $\bar q_{V1} q_{V2}$.
I call these $M_{12}$ and $f_{12}$, respectively.
Note that this state is a flavor non-singlet, 
and so there are no disconnected contributions to its propagator.
Various limits of the general result are of interest:
\begin{itemize}
\item
If $m_1=m_2$ one obtains a non-singlet
pion composed of degenerate valence quarks. 
I refer to the results in this limit as $M_{11}$ and $f_{11}$,
or generically as $M_{VV}$ and $f_{VV}$.
\item
Setting $m_2=m_S$ one obtains a pion with only one quenched quark.
I refer to the results for this pion as $M_{1S}$ and $f_{1S}$,
or generically as $M_{VS}$ and $f_{VS}$.
\item
The case $m_1=m_2=m_S$ is special, for then one is considering the physical
unquenched pion in which the valence and sea quarks have the same mass.%
\footnote{This is only true for $N\ge2$. One cannot make a non-singlet pion 
if there is only a single flavor of sea quark. Thus if $N=1$, the results
for $M_{12}$ in the limit that $m_1=m_2=m_S$ do not correspond to those
for an unquenched pion.}
This case deserves a separate notation, and I follow Ref.~\cite{SESAM}
by denoting its mass and decay constant $M_{SS}$ and $f_{SS}$, respectively.
\end{itemize}
There is no problem in principle 
extending the calculation to pseudoscalars which have a 
flavor singlet component.
I have not done so for two reasons.
First, the results involve a number of unknown constants not present
in the expressions for the non-singlet masses
(an example is given in Ref.~\cite{BGPQ}).
Second, nearly all simulations calculate only non-singlet masses,
because the annihilation diagrams needed for singlet states require 
much greater computational resources.

The leading order results for non-singlet pions are
\be
\left[M_{12}^2\right]_{\rm tree} = \mu (m_1 + m_2)\,, \qquad
\left[f_{12}\right]_{\rm tree} = f \,.
\ee
The result for $M_{12}$ agrees with that given 
in the Introduction, Eq.~(\ref{eq:stagg}).
In particular, $M_{12}$ is, at leading order, independent of 
the sea quark mass $m_S$.

I have calculated the one-loop results using
dimensional regularization and $\overline{MS}$ subtraction.
To present these I use the notation that 
$y_{12} = \mu (m_1+m_2)/\Lambda_\chi^2$, $y_{SS}= 2 \mu m_S/\Lambda_\chi^2$,
etc., where $\Lambda_\chi=4\pi f$.
The result for the non-leading contribution to the pion mass is
\bea
\left[ M_{12}^2\right]_{\rm 1-loop} &=& {\mu(m_1+m_2)} \left\{
{1\over N} \left[
y_{11} (y_{SS} - y_{11}) \ln y_{11} -
y_{22} (y_{SS} - y_{22}) \ln y_{22}  
\over
y_{22} - y_{11} \right] 
\right.\nonumber \\ 
&&\mbox{}\qquad\qquad\qquad
 \left. \vphantom{\left[ {y_{22}\over y_{11}}\right]}
+  y_{12} (2 \alpha_8 - \alpha_5)
+  y_{SS} N (2 \alpha_6-\alpha_4) \right\}
\,.  \label{eq:M12res}
\eea
The constants $\alpha_i$ are to be evaluated at the scale $\Lambda_\chi$.
In the degenerate limit this becomes
\bea
\left[ M_{VV}^2\right]_{\rm 1-loop} &=& {2 \mu m_V} \left\{
\frac{1}{N} \left[
(2 y_{VV} - y_{SS}) \ln y_{VV} + (y_{VV} - y_{SS}) \right] \right.
\nonumber \\
&&\mbox{}\qquad\quad \left. \vphantom{\frac{1}{N}}
+  y_{VV} (2 \alpha_8 - \alpha_5)
+  y_{SS} N (2 \alpha_6-\alpha_4) \right\}
\,. \label{eq:MVVres}
\eea
If we further set $m_V=m_S$ we obtain
\be
\left[ M_{SS}^2\right]_{\rm 1-loop} = {2 \mu m_S} \left\{
{1\over N} y_{SS} \ln y_{SS} 
+  y_{SS} \left[ (2 \alpha_8 - \alpha_5) +
N (2 \alpha_6-\alpha_4) \right] \right\}
\,. \label{eq:MSSres}
\ee
This agrees with the result from standard chiral perturbation 
theory~\cite{GL}.

The enhanced chiral logarithmic corrections 
discussed in the Introduction appear in the result
for $M_{VV}^2$, Eq.~(\ref{eq:MVVres}):
these are corrections of relative size $m_{S} \ln m_{V}$. 
Enhanced corrections are present also in the general result
Eq.~(\ref{eq:M12res}) if we take $m_1$ and $m_2$ to zero in fixed ratio.
They are absent, however, from the mass of the pion composed of one
valence quark and one sea quark. This is obtained by setting $m_1=m_V$
and $m_2=m_S$, yielding
\be
\left[ M_{VS}^2\right]_{\rm 1-loop} = {\mu (m_V+m_S)} \left\{
{1\over N} y_{VV} \ln y_{VV} 
+  y_{VS}  (2 \alpha_8 - \alpha_5)
+  y_{SS} N (2 \alpha_6-\alpha_4) \right\}
\,. \label{eq:MVSres}
\ee
The chiral correction here is of relative size $m_{V} \ln m_{V}$,
and thus vanishes when $m_V\to0$.

The fact that all the loop terms are proportional to $1/N$ appears
to be an accident. One might have expected terms proportional to $N$,
since there are $N$ mesons of the form $\bar q_V q_S$ which can appear
in loops. It turns out, however, that such contributions 
cancel in the final result. What remains is the contribution
from loops involving ``hairpin'' vertices. \\

The corresponding results for decay constants are
\bea
{f_{12} \over f} &=& 1 
- {N \over 4} (y_{1S} \ln y_{1S} + y_{2S} \ln y_{2S})
\nonumber\\ &&\mbox{}\qquad
+ {1 \over 2N} \left(
{y_{11} y_{22} - y_{SS} y_{12} \over y_{22}-y_{11}}
\ln {y_{11}\over y_{22}}
+ y_{12}-y_{SS} \right)
\nonumber\\
&&\mbox{}\qquad + \frac12 \alpha_5 y_{12} + \frac12 \alpha_4  N y_{SS}
\,,\label{eq:f12res}\\
{f_{VV} \over f} &=& 1 - {N\over2} y_{VS} \ln y_{VS}
+ \frac12\alpha_5 y_{VV} + \frac12 \alpha_4 N y_{SS}
\,,\label{eq:fVVres}\\
{f_{VS} \over f} &=& 1 
- {N\over4}   (y_{VS} \ln y_{VS} + y_{SS} \ln y_{SS})
- {1\over 4N}  \left( 
y_{SS} \ln{y_{VV}\over y_{SS}} + y_{SS}-y_{VV} \right)
\nonumber\\&&\mbox{}\qquad
+ \frac12\alpha_5 y_{VS} + \frac12 \alpha_4 N y_{SS}
\,.\label{eq:fVSres}
\eea
Comparing these with the results for masses,
we see that the enhanced chiral logarithms 
survive here in $f_{VS}$ but not in $f_{VV}$,
which is the opposite of the situation for the masses. 
For both quantities the enhanced logarithms are multiplied by $1/N$.
The decay constants do, however, have contributions proportional
to $N$, but from logarithms which are not enhanced. \\

To give a sense of the size and form of the corrections, 
I display the results for the $VV$, $VS$ and $SS$ pions for $N=2$.
To convert meson masses to physical units I take $f= 0.1\GeV$
(which is the approximate value for this constant in QCD).
To convert quark masses to units of the physical strange quark mass, $\ms$,
I assume the leading order result $\mu \ms =M_{K,\rm phys}^2$. 
This ignores the shift due to the one-loop corrections, 
but, as we will see, these corrections are of moderate size.
Finally, I set the analytic constants, $\alpha_{4-8}$, to zero.
This is the simplest choice given that we do not 
know what values to use for $N=2$. I have checked that the essential
features of the plots are unchanged if $\alpha_{4-8}$ are set to
the values they take in QCD.

\begin{figure}[t]
\vspace{-0.1truein}
  \centerline{\psfig{file=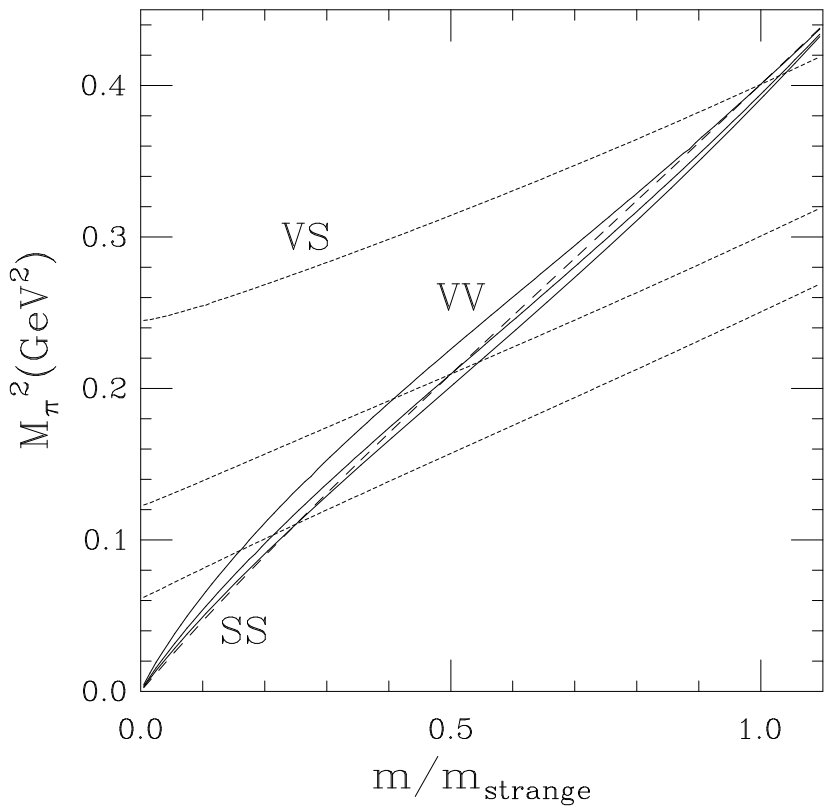,height=5truein}}
\vspace{-0.3truein}
\caption{Predictions for pion masses using values for the
parameters discussed in the text. The solid and short-dashed
curves show $M_{VV}^2$ and $M_{VS}^2$,
including one-loop contributions, plotted against $m_V$.
The three sets of curves correspond to $m_S=\ms$, $\ms/2$, and $\ms/4$
as one moves from top to bottom.
The long-dashed curve is the result for $M_{SS}^2$ 
at one-loop plotted against $m_S$.}
\label{fig:mpi}
\end{figure}

Figure~\ref{fig:mpi} shows the one-loop predictions for the masses of
the ${VV}$ and ${VS}$ pions plotted against the valence quark mass $m_V$, 
for three values of the sea quark mass, $m_S=\ms/4$, $\ms/2$ and $\ms$.
This range is chosen to cover the typical values for ``light'' quarks
used in present simulations.
The chiral expansion is likely to break down at the upper end of this range.
By construction, the $VV$ and $VS$ curves must cross when $m_V=m_S$.
For purposes of comparison, I have also included the one-loop
result for $M_{SS}^2$ plotted against $m_S$.

Lowest order chiral perturbation theory predicts that all curves are
linear, with the three for $M_{VV}$ and that for $M_{SS}$ coinciding,
while the curves for $M_{VS}$ have half the slope of those for $M_{VV}$.
We see that, although one-loop corrections do change this prediction, 
the major features of the leading order result remain.
The most significant change is that $M_{VV}$ and $M_{SS}$ no longer
coincide, with the curves for $M_{VV}$ showing some curvature.

\begin{figure}[tb]
\vspace{-0.1truein}
  \centerline{\psfig{file=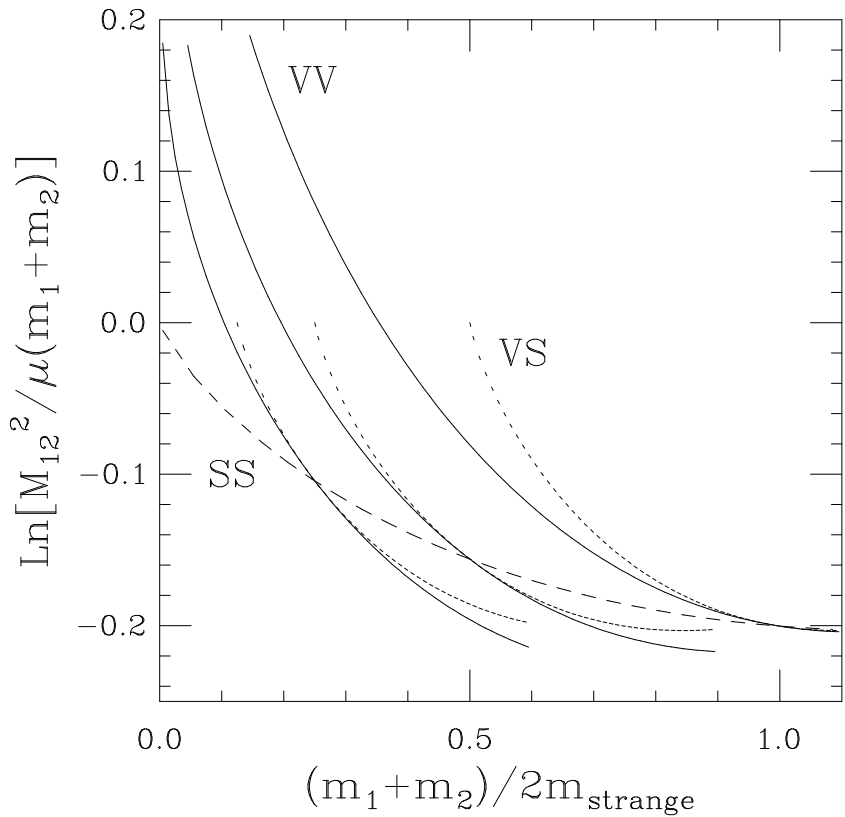,height=5truein}}
\vspace{-0.3truein}
\caption{Predictions for $\ln M_{VV}^2/2 \mu m_V$, 
$\ln M_{VS}^2/\mu(m_V+m_S)$ and $\ln M_{SS}^2/2\mu m_S$,
including one-loop contributions,
plotted against the average quark mass. Notation as in Fig.~\ref{fig:mpi}.
The three sets of curves for ${VV}$ and ${VS}$ mesons
correspond to $m_S=\ms/4$, $\ms/2$ and $\ms$ as one moves from left to right.}
\label{fig:logmpi}
\end{figure}

To magnify the difference between $M_{VV}$ and $M_{SS}$,
it is advantageous to consider quantities in which the leading
order quark mass dependence has been removed.
One such quantity, $\ln[M_{12}^2/\mu(m_1+m_2)]$,
is plotted in Fig.~\ref{fig:logmpi}.
The enhanced chiral logarithm causes the $VV$ curves
to diverge as $m_V\to0$.
A similar divergence is predicted for quenched QCD, and plots of this kind
are a useful way of searching for this divergence.
Such a search will not, however, be easy.
For one thing, it is hard to distinguish 
logarithms from the linear dependence predicted by analytic terms 
unless one has a large range of quark masses.\footnote{%
One cannot overcome this problem by considering the difference between the
$VV$ and $SS$ curves---this cancels the analytic term proportional
to $2 \alpha_8 - \alpha_5$, but leaves the term proportional
to $2 \alpha_6-\alpha_4$.}
It is also difficult to separate logarithms from the
$1/m_V$ behavior associated with finite volume 
effects~\cite{sslat96,mawhinney}.

An alternative approach is to consider quantities in which
the analytic terms cancel. One example is the difference between
the masses of $VV$ and $VS$ mesons composed of quarks having
the same average mass evaluated at the same sea quark mass.
This vanishes at tree level, but not at one-loop:
\be
M_{11}^2(m_1,m_S)-M_{2S}^2(m_2=2m_1-m_S,m_S) =
2 \mu m_1 \frac{1}{N}
\left\{ y_{22} \ln {y_{11} \over y_{22}} + y_{22}-y_{11} \right\}
\label{eq:VVmVS} \,.
\ee
This may also be a more practical difference to study in detail as it
relies only on changing the valence quark mass.
Once one has determined $f$, this difference is, for small enough
quark mass, a prediction of partially quenched chiral perturbation theory
free of unknown parameters.
To illustrate this prediction 
I have included the results for the ${VS}$ mesons in Fig.~\ref{fig:logmpi},
but now plotted against the average quark mass $(m_V+m_S)/2$ rather
than $m_V$. The quantity in Eq.~(\ref{eq:VVmVS}) is simply
the difference between the $VV$ and $VS$ curves at fixed $m_S$.
This brings out a striking (and seemingly accidental)
prediction of chiral perturbation theory: the curves for
$M_{VV}^2$ and $M_{VS}^2$ have the same derivative at $m_V=m_S$.
This is true at leading order and is not affected by loop corrections.
Because of this, the predicted difference between the $VV$ and $VS$ curves
is small (and gets smaller as the quark masses decrease).

One can also consider a generalization of this difference
which allows more flexibility in
testing the predictions of chiral perturbation theory. 
Imagine working at fixed $m_S$, 
and pick three valence quark masses which satisfy
\be
m_1 + m_3 = 2 m_2 \,.
\ee
The quantity of interest is the relative difference between the masses
of the ``$13$'' and ``$22$'' mesons. This vanishes at tree-level,
but at one-loop takes the form
\be
{M_{13}^2 - M_{22}^2 \over M_{13}^2 + M_{22}^2}
= {1 \over 2 N} \left\{
{y_{11} (y_{SS}-y_{11}) \ln(y_{11}/y_{22}) \over y_{33}-y_{11}}
- {y_{11}-y_{SS} \over 2} + (1\leftrightarrow 3) \right\}
\,.
\label{eq:massdiff}
\ee
In evaluating the r.h.s. of this expression  it is legitimate to use the
pion masses themselves, rather than their lowest order expression in
terms of quark masses, since the difference is a two-loop effect.
The same is true for Eq.~(\ref{eq:VVmVS}).

\begin{figure}[tb]
\vspace{-0.1truein}
  \centerline{\psfig{file=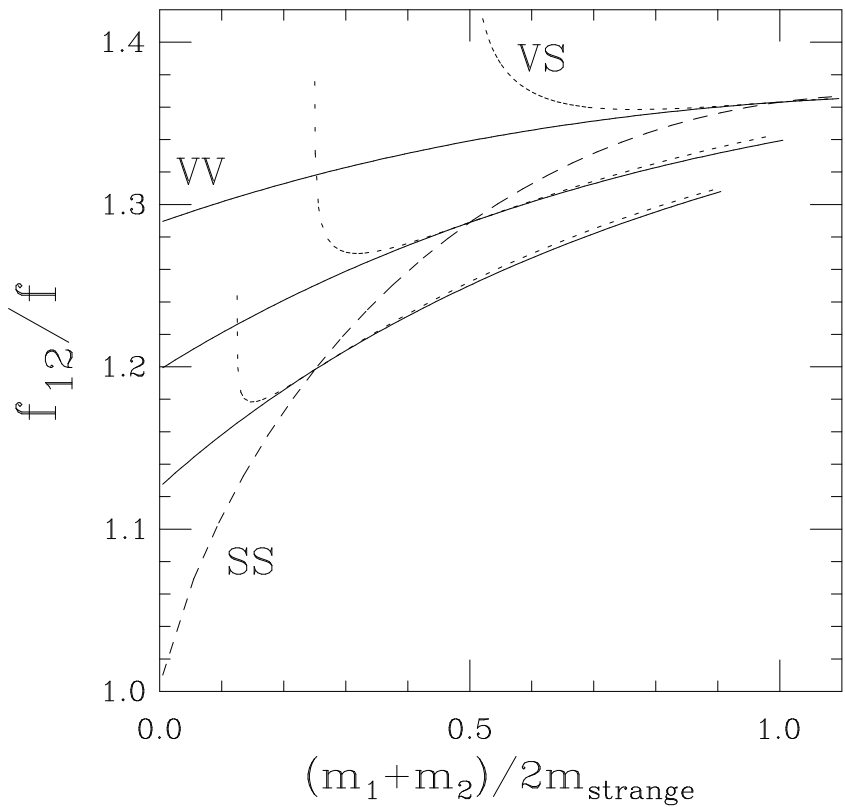,height=5truein}}
\vspace{-0.3truein}
\caption{Results for decay constants plotted against average quark
mass. Notation as in Fig.~\ref{fig:mpi}. 
As one moves from top to bottom,
the three sets of curves correspond to $m_S=\ms$, $\ms/2$, and $\ms/4$.}

\label{fig:f}
\end{figure}

The results for decay constants are shown in Fig.~\ref{fig:f}.
I have plotted them against the average quark mass so that
the analytic contributions cancel in the difference between $f_{VV}$ 
and $f_{VS}$ at fixed $m_S$.\footnote{%
It should be borne in mind, however, that the detailed shape of the
curves, and the difference between the VV and SS results, does depend
on the choice of $\alpha_4$ and $\alpha_5$.}
It turns out that, as for pion masses,
the $VV$ and $VS$ curves have the same derivative at $m_V=m_S$.
Note that the corrections are of moderate size even for $m_V,m_S\approx \ms$.
The only exception is that $f_{VS}$ diverges in the limit $m_V\to 0$
due to the enhanced chiral logarithm.

One way of looking for this enhanced logarithm is to form the ratio
\be
R_{BG} = {f_{12} \over \sqrt{f_{11} f_{22} }}
\ee
at fixed $m_S$.
This quantity was introduced by Bernard and Golterman as a way of testing
chiral perturbation theory for quenched QCD \cite{BG}, but turns
out also to be useful in partially quenched theories.
The analytic contributions cancel in $R_{BG}$, and one finds
\be
R_{BG}-1 = {1\over 2 N} \left\{
\left[{y_{11} y_{22} \over y_{22}-y_{11} }\ln {y_{11}\over y_{22}}
      + y_{12} \right]
- y_{SS} \left[{y_{12} \over y_{22}-y_{11}}\ln {y_{11}\over y_{22}}
      + 1 \right] \right\}
\,.
\label{eq:RBGres}
\ee
It is again legitimate to use the actual pion masses when evaluating
this result. Note that the prediction for this quantity diverges
as $m_1\to0$ with fixed $m_2$ and $m_s$.
A similar divergence is predicted for quenched QCD, 
with $y_{SS}^2/ 2N$ replaced by $m_0^2/3\Lambda_\chi^2$ \cite{BG}.

\section{Baryon masses}\label{sec:baryon}

The enhanced chiral loops which lead to the divergences also appear
in other quantities. In particular, one can use partially quenched
chiral perturbation theory to study the behavior of baryon masses,
using a straightforward extension of the methods developed for
quenched baryons \cite{LS}.
I have not carried out a detailed calculation, but
it is easy to determine the general form of the dependence on the
quark masses. The result for baryons
composed of three degenerate valence quarks is
\be
M_{VVV} = M_0 + c_1 M_{VV} M_{SS}^2 + c_2 M_{VV}^2 + c_3 M_{SS}^2
+ O(M_{VV}^3) 
\,.
\ee
This form applies for both spin 1/2 and 3/2 baryons,
although the coefficients (including $M_0$) depend on the spin.
The coefficient $c_1$ can in principle be predicted in terms of the
pion-nucleon couplings $F$ and $D$, and the decay constant $f$. 
The other coefficients are $N-$dependent unknown constants.
The term proportional to $c_1$ is the analogue of the
enhanced chiral logarithms found above. It does not diverge as
$m_V\to0$, but it is the dominant correction for sufficiently small
$m_V$ at fixed $m_S$. 
This is no longer true in the unquenched limit, $m_V=m_S$, 
for then the $c_1$ term is of $O(M_{SS}^3)$.

\section{Conclusions}\label{sec:conc}

Partially quenched theories are a step on the way from quenched
to full QCD.
They allow one to partially probe the dynamics of light quarks by sending
the valence quark mass towards zero, while holding the sea quark mass fixed. 
In this paper I have investigated the errors that this procedure introduces.
The situation turns out be rather subtle in that the errors only show
up at non-leading order, but they nevertheless diverge (in relative size)
as $m_V\to 0$.
I have suggested a number of combinations of pion masses and
decay constants with which to search for such divergences.

The entire discussion has assumed that we know what the lattice
quark masses are, as is the case for staggered fermions.
As mentioned in the Introduction, with Wilson fermions
there are problems in determining the quark mass from the hopping parameter
in partially quenched theories.
These problems can be avoided, however, by determining the
quark mass non-perturbatively using the PCAC equation. 
In this way the predictions can be tested for Wilson fermions.
The only disadvantage compared to staggered fermions is that the predictions
will hold up to discretization errors of $O(a)$ rather than $O(a^2)$.
Even without a non-perturbative determination of the quark masses,
Eqs.~(\ref{eq:VVmVS}), 
(\ref{eq:massdiff}) and (\ref{eq:RBGres}) can still be tested
with Wilson fermions.

\section*{Acknowledgements}
I thank Tanmoy Bhattacharya and Rajan Gupta for discussions.
This work was supported in part
by the U.S. Department of Energy grant DE-FG03-96ER40956.

\end{document}